\documentstyle[preprint,prl,aps]{revtex}
\begin{document}
\def\sn2{$\sin^22\theta$}
\def\dm2{$\Delta m^2$}
\def\ch2{$\chi^2$}
\draft
\begin{titlepage}
\preprint{\vbox{\baselineskip 15pt{
\hbox{Ref. SISSA 41/94/EP}
\hbox{FERMILAB -- PUB -- 94/188--T}
\hbox{IFP -- 295 -- UNC}
\hbox{hep-ph/9408234}
\hbox{July 1994}
\hbox{}}}}
\vskip -0.2cm
\title{ \large \bf TESTING THE VACUUM OSCILLATION AND THE MSW SOLUTIONS OF
THE SOLAR NEUTRINO PROBLEM}
\author{P. I. Krastev\footnote{Permanent address:
Institute of Nuclear Research and Nuclear Energy, Bulgarian Academy
of Sciences, BG--1784 Sofia, Bulgaria.}}
\address{Institute of
Field Physics, Department of Physics and Astronomy, \\
The University
of North Carolina, Chapel Hill, NC 27599-3255,}
\vglue 0.2cm
\author{S.T. Petcov$^*$}
\address{Scuola Internazionale Superiore di Studi Avanzati, and Istituto\\
Nazionale di Fizica Nucleare, Sezione di Trieste, I-34013 Trieste, Italy}
\maketitle
\begin{abstract}
\begin{minipage}{5in}
\baselineskip 16pt
Solar model independent tests of the vacuum oscillation and MSW
solutions of the solar neutrino problem are considered. Detailed
predictions for time (seasonal) variations of the signals in the
future solar neutrino detectors (SNO, Super Kamiokande, BOREXINO,
HELLAZ), if solar neutrinos take part in vacuum oscillations, are
given. Results on the distortions of the spectra of $^{8}$B neutrinos,
and of e$^{-}$ from the reaction $\nu + e^{-} \rightarrow \nu + e^{-}$
induced by $^{8}$B neutrinos, in the cases of vacuum oscillations or
MSW transitions are presented for a large number of values of the
relevant parameters.  The possibilities to distinguish between the
vacuum oscillation, the MSW adiabatic, and the MSW nonadiabatic
transitions (solutions) in the future solar neutrino experiments are
discussed.
\end{minipage}
\end{abstract}
\end{titlepage}
\newpage

\hsize 16.5truecm
\vsize 24.0truecm
\def\dm{$\Delta m^2$\hskip 0.1cm }
\def\dmf{\Delta m^2}
\def\sn{$\sin^2 2\theta$\hskip 0.1cm }
\def\snf{\sin^2 2\theta}
\def\trna{$\nu_e \rightarrow \nu_a$}
\def\trnm{$\nu_e \rightarrow \nu_{\mu}$}
\def\trns{$\nu_e \leftrightarrow \nu_s$}
\def\trnat{$\nu_e \leftrightarrow \nu_a$}
\def\trnmt{$\nu_e \leftrightarrow \nu_{\mu}$}
\def\trne{$\nu_e \rightarrow \nu_e$}
\def\trnst{$\nu_e \leftrightarrow \nu_s$}
\font\eightrm=cmr8
\def\aprle{\buildrel < \over {_{\sim}}}
\def\aprge{\buildrel > \over {_{\sim}}}
\renewcommand{\thefootnote}{\arabic{footnote}}
\setcounter{footnote}{0}
\vskip 0.0cm
\leftline{\bf 1. INTRODUCTION}
\vskip 0.4cm
\indent With the publication in 1991 and 1992 of the results of the
Ga-Ge solar neutrino experiments [1,2] it became clear that the data
from the first generation of solar neutrino detectors will not be
sufficient to resolve the solar neutrino problem [3-6] which has been
with us for more than 20 years. If the latest data provided by the
pioneer Davis et al.  [3], Kamiokande [7], SAGE [1] and GALLEX [2]
experiments are correct, an astrophysical explanation of the solar
neutrino deficit seems unlikely at present [8] (for a recent
discussion see [9] and [10]). At the same time, the current solar
neutrino observations admit several rather different neutrino physics
interpretations which require the existence of unconventional neutrino
intrinsic properties (mass, mixing, magnetic moment) and/or couplings
(e.g., flavour changing neutral current (FCNC) interactions). These
include: i) oscillations in vacuum [11] of the solar $\nu_e$ into
different weak eigenstate neutrinos ($\nu_{\mu}$ and/or $\nu_{\tau}$,
and/or sterile neutrinos, $\nu_s$) on the way from the surface of the
Sun to the Earth [12] \footnote{For earlier discussions see, e.g., the
references quoted in [12].}, ii) MSW transitions [13]
$\nu_e\rightarrow\nu_{\mu(\tau)}$, and/or $\nu_e\rightarrow\nu_s$,
while the solar neutrinos propagate from the central part to the
surface of the Sun [14] \footnote {The MSW solution has been studied
by many authors before and after the publication of the results of the
two Ga-Ge experiments: see, e.g., refs. [2,15] and the references
quoted in [2,14,15].}, iii) solar $\nu_e$ resonant spin or
spin-flavour precession (RSFP) [16] in the magnetic field of the Sun
[17], and iv) matter-enhanced transitions, for instance
$\nu_e\rightarrow\nu_{\tau}$, in the Sun, induced by FCNC interactions
of the solar $\nu_e$ with the particles forming the solar matter
[18,19] (these transitions can take place even in the case of absence
of lepton mixing in vacuum and massless neutrinos [18])
\footnote{The solar neutrino decay
hypothesis [20] is disfavoured [21], while mechanisms leading to
universal suppression of the fluxes of $^{8}$B, $^{7}$Be, pp,
etc. neutrinos due to $\nu_e \rightarrow \nu_{s}$ transitions are
ruled out, by the current solar neutrino data, if one uses the
standard solar model (SSM) predictions of refs. [5,6] in the relevant
analyses.}.  Although the experiments of Davis et al., Kamiokande,
SAGE and GALLEX will continue to run for at least few more years and
the accuracy of the data they provided will improve, no substantial
changes of the latter are expected \footnote {A priori, one cannot
totally rule out the possibility of surprises in the next few
years. The planned calibrations of the GALLEX and SAGE detectors will
be crucial for the conclusive determination of the characteristics of
the solar neutrino flux inferred from the current data.}  and no
qualitatively new data will be available before solar neutrino
detectors of the second generation - SNO [22], Super Kamiokande [23],
BOREXINO [24], ICARUS [25], and HELLAZ [26], become operational in the
second half of the 90-ies \footnote{Two of these detectors - SNO and
Super Kamiokande, are under construction, BOREXINO and ICARUS are at
the stage of prototype construction and/or testing, and the
possibility to build HELLAZ is being studied.}. Our hopes for finding
the cause of the solar neutrino deficit and for getting more precise
information about the physical conditions in the central part of the
Sun, where the neutrinos are being produced, are now associated with
these future experiments.

  In the present article we continue the studies [14,27] of the
possible solar model independent tests of the vacuum oscillation and
the MSW solutions of the solar neutrino problem. The importance of
these tests is difficult to overestimate given the fact that the solar
model predictions for the $^{8}$B neutrino flux may have rather large
uncertainties. We present results on the specific seasonal time
variations of the signals in the future solar neutrino experiments
(SNO, Super Kamiokande, BOREXINO, HELLAZ), predicted if the solar
neutrino deficit is caused by vacuum oscillations of solar
neutrinos. We give also detailed predictions for the distortions of
the spectra of $^{8}$B neutrinos, and of e$^{-}$ from the reaction
$\nu + e^{-} \rightarrow \nu + e^{-}$ caused by the $^{8}$B neutrinos,
in the cases of the vacuum oscillation and the MSW solutions. Neither
seasonal time variations (apart from the standard $\sim~$7\%
geometrical one), nor substantial distortions of the spectra of the
$^{8}$B, pp and CNO neutrinos (greater than $\sim 10^{-3}~$E, E being
the neutrino energy) are predicted to arise due to the specific
physical conditions in the interior of the Sun [28].  The
possibilities to distinguish between the vacuum oscillation, the MSW
adiabatic, and the MSW nonadiabatic solutions of the solar neutrino
problem using the data from the future SNO and Super Kamiokande
experiments are discussed \footnote {For alternative solar model
independent tests of these solutions, based on the future SNO and
Super Kamiokande data see refs. [29].}.  Updated results on the MSW
solution of the solar neutrino problem are also given.
\vskip 0.4cm
\leftline{\bf 2. VACUUM OSCILLATIONS OF SOLAR NEUTRINOS: PREDICTED}
{\bf SEASONAL VARIATION EFFECTS AND SPECTRA DISTORTIONS}
\vskip 0.3cm
\indent  The two--neutrino vacuum oscillation solution of the solar neutrino
problem has been re-analyzed recently \cite{KP4} using the latest data
from all currently operating neutrino experiments (Homestake,
Kamiokande III, GALLEX and SAGE). The analysis was based on the
predictions of the solar model of Bahcall and Pinsonneault
\cite{BahP}. It was found that the two--neutrino oscillations
involving the $\nu_e$ and an active neutrino,
$\nu_e\leftrightarrow\nu_{\mu (\tau)}$, provide a not very good (but
acceptable) quality of the $\chi^2$--fit to the mean event rate solar
neutrino data, while the oscillations into sterile neutrino $\nu_s$,
$\nu_e\leftrightarrow\nu_s$, give a poor fit of the data: the
$\nu_e\leftrightarrow\nu_{\mu (\tau)}$ oscillations are ruled out at
90 \% C.L., but are acceptable at 95 \% C.L., while the
$\nu_e\leftrightarrow\nu_s$ oscillations are ruled out at 99 \%
C.L. The results are rather different if one uses the data available
from each particular run of measurement of the Homestake, Kamiokande
II, GALLEX and SAGE collaboration in the $\chi^2$--analysis (for
details see ref. \cite{KP4}). Both the $\nu_e\leftrightarrow\nu_{\mu
(\tau)}$ and $\nu_e\leftrightarrow\nu_s$ oscillation hypotheses give
good fits to the run--by--run solar neutrino data, being acceptable
even at 68 \% C.L. The regions of values of the two parameters,
$\Delta m^2$ and $\sin^22\theta$, characterizing the two--neutrino
oscillations of the solar $\nu_e$, which are allowed (at 90 \% C.L.)
by the run--by--run data, lie in the following narrow intervals
\cite{KP4}:
$$\nu_e\leftrightarrow\nu_{\mu (\tau)}:~~~~~~~~ 5.7\times10^{-11} {\rm
eV}^2\stackrel{<}{\sim}\Delta m^2 \stackrel{<}{\sim} 1.1\times
10^{-10} {\rm eV}^2, \eqno(1a)$$ $$\sin^22\theta \stackrel{>}{\sim}
0.75,\eqno(1b)$$
\noindent and
$$\nu_e\leftrightarrow\nu_s:~~~~~~~~ 5.0\times10^{-11} {\rm
eV}^2\stackrel{<}{\sim}\Delta m^2 \stackrel{<}{\sim} 6.6\times
10^{-11} {\rm eV}^2,\eqno(2a)$$ $$\sin^22\theta \stackrel{>}{\sim}
0.8.\eqno(2b)$$
\vskip 0.4cm
\leftline{\bf 2.1 Seasonal Variations of Signals}
\vskip 0.2cm
The probability that a solar electron neutrino with energy E will not
change into $\nu_{\mu (\tau)}$ (or $\nu_s$) on its way to the Earth
when $\nu_e\leftrightarrow\nu_{\mu (\tau)}$
($\nu_e\leftrightarrow\nu_s$) oscillations take place, has the form:
\vskip -0.4truecm
$${\rm P(\nu_e\rightarrow\nu_e; R(t), E)} = 1 - {1\over 2}
\sin^22\theta~\bigl[ 1 - \cos 2\pi {\rm R(t)\over
L_v}\bigr],\eqno(3)$$
\noindent where ${\rm L_v = 4\pi E}/\Delta m^2$ is the oscillation length in
vacuum,
$${\rm R(t) = R_0~\bigl[ 1 - \epsilon\cos 2\pi {t\over T}}\bigr],\eqno(4)$$
\noindent is the Sun--Earth distance at time t of the year (T = 365 days),
${\rm R_0} = 1.496\times 10^8$ km and $\epsilon = 0.0167$ being the mean
Sun--Earth distance and the ellipticity of the Earth orbit around the Sun.

   For E $\simeq$ 1 MeV and the values of $\Delta m^2$ from the
intervals (1a) and (1b) one has: ${\rm L_v \simeq 2\pi(2\epsilon
R_0)}$, where $2\epsilon {\rm R_0}$ is the variation of the Sun--Earth
distance in the period December -- June. This implies that if solar
neutrinos take part in vacuum oscillations, the flux of solar
neutrinos will exhibit seasonal variations \footnote{The possibility
of seasonal variations of the flux of solar neutrinos when the latter
take part in oscillations in vacuum was indicated first by
B.Ya. Pomeranchuk (see, e.g., ref. \cite{BilP}).}. The magnitude of the
time variations depends, in particular, on the energy of solar
neutrinos and will be different for the $^8$B, $^7$Be, pp, pep and the
CNO neutrinos. Obviously, if the integration over the neutrino energy
renders the oscillating term in the expression for ${\rm
P(\nu_e\rightarrow\nu_e; R(t), E)}$ negligible (as is the case of pp
neutrinos \cite{KP1}), the energy integrated observables will not
exhibit seasonal (time) variations.

   We have depicted in Figs. 1 and 2 the expected time variation of
the ratio of the predicted signals (event rates) due to the $^{8}$B,
pp, $^{7}$Be and pep neutrinos in the case of (two--neutrino)
oscillations of solar neutrinos, to the ones in the absence of
oscillations. The results shown in these figures are valid for
experiments detecting the solar neutrinos via the $\nu-{\rm e}^{-}$
elastic scattering reaction (Super Kamiokande, BOREXINO, HELLAZ,
etc.). The electron kinetic energy detection threshold was taken in
the calculations of the signals due to the $^{8}$B, the pp, the
$^{7}$Be and pep neutrinos to be 5 MeV, 0.1 MeV, and 0 MeV,
respectively; the possible effects of the detectors efficiencies and
finite electron energy resolution were not included in the
calculations (we leave it to our colleagues--experimentalists to take
into account these effects in accordance with the specific
characteristics of their respective detectors). Electrons with kinetic
energy T$_e \aprge~$5 MeV will be detected in the Super Kamiokande
experiment. The HELLAZ detector is planned to be sensitive to e$^{-}$
(from the pp neutrino induced reaction) with T$_{e}~\aprge~$0.1
MeV. As for the BOREXINO detector in which $\sim$ 90\% of the event
rate is predicted to be generated by the 0.862 MeV $^{7}$Be neutrinos,
it is expected that the signal to background ratio will allow one to
extract the $^{7}$Be neutrino signal for e$^{-}$ with kinetic energy
in the interval 0.25 MeV$~\aprle~$ T$_{e}~\aprle~$0.66 MeV. We have
checked that reducing the interval 0$~\leq~$T$_{e}~\aprle~$0.66 MeV
used in our calculations to the one reflecting the currently envisaged
detection capabilities of BOREXINO has no observable effect on the
results (the two $^{7}$Be(862) curves, corresponding to the two
different intervals of integration would be indistinguishable if both
were plotted in Figs. 1 and 2).

     In the calculations of the signals (event rates) in the case of
vacuum oscillations we have taken into account also the standard
R$^{-2}$(t) dependence of the values of the different solar neutrino
flux components at the Earth surface. Finally, the SSM predicted event
rates (signals) have been obtained by dividing by T the one year total
number of events calculated within the SSM [5] (assuming 100\%
detection efficiency). Thus, in the absence of vacuum oscillations the
plotted (theoretical) ratios will change from $(1 - \epsilon)^{-2}$ in
December to $(1 + \epsilon)^{-2}$ in June, while the ratio of measured
to the SSM predicted [5] signals will vary from A$(1 - \epsilon)^{-2}$
to A$(1 + \epsilon)^{-2}$, where A is a constant which can be different
for the different (pp, $^{7}$Be, $^{8}$B, and pep neutrino induced)
signals (A = 1 if the SSM prediction [5] for the flux of the
corresponding neutrinos (pp, and/or $^{7}$Be, and/or $^{8}$B, and/or
pep) is correct).

   The results shown in Figs. 1 and Figs. 2a--2d correspond to solar
$\nu_e$ oscillations into active neutrino,
$\nu_e\leftrightarrow\nu_{\mu (\tau)}$, while in Fig. 2e and Fig. 2f
we have depicted results in the case of oscillations into sterile
neutrino, $\nu_e\leftrightarrow\nu_s$.

    As was indicated on the basis of few numerical examples in
[14,27], the most dramatic seasonal variations are predicted to be
exhibited by the signals due to the monochromatic $^7$Be and pep
neutrinos. Typically, the differences between the signals in December
and June are the largest. However, as was noted in \cite{KP2}, for
certain values of the parameter $\Delta m^2$ the signals in December
(or June) and March (or September) differ most (see, e.g., Figs. 1c,
1d and 2f). Figs. 1 and 2 demonstrate also that the predicted
magnitude and explicit form of the time variations of the $^7$Be and
pep neutrino induced event rates in the $\nu-e^-$ elastic scattering
experiments are extremely sensitive to the value of $\Delta m^2$ and
change drastically even for relatively small variations of this
parameter.

   Since the pp neutrinos have a rather low energy (E $\leq 0.42$
MeV), for most of the values of $\Delta m^2$ from the intervals (1a)
and (2a) the inequality $2\pi R_o >> L_v$ holds. As a consequence, the
integration over the recoil e$^-$ energy in the calculations of the pp
neutrino induced signals renders the oscillation term in the
probability (3) negligible \cite{KP1}. Therefore the predicted
seasonal change of the energy integrated signals due to the pp
neutrinos in the $\nu~-~e^-$ scattering experiments coincides with the
standard 7\% geometrical one (Figs. 1, 2a--2c, and 2e), except for
values of $\Delta m^2 \approx 5\times10^{-11}~eV^2$, for which
the vacuum oscillations lead to rather small deviations from it
(Figs. 2c and 2f).

  The seasonal changes of the signal due to the $^8$B neutrinos do not
exceed approximately 15\% \cite{KP2}. Such variations are not
detectable in the currently running experiments. However, the high
statistics future solar neutrino experiments Super Kamiokande, SNO,
and ICARUS are envisaged to accumulate (between 3000 and 4000 events
per year) will allow to detect even rather small (few percent)
differences between the signals in December and June.  For Super
Kamiokande (ICARUS) and SNO detectors the effect of time variations is
shown separately in Figs. 3a (3b), 3d (3e), and in Figs. 3c and 3f,
respectively, using a different normalization of the signals and a
proper scale. The predictions for the Super Kamiokande detector
depicted in Figs. 3a (3d) and 3b (3e) differ in the value of the
neutrino threshold energy, E$_{th}$, used in the calculations (see
further): E$_{th} = 5~$ MeV and E$_{th} = 7.5~$ MeV, respectively. The
smaller magnitude of the effect in comparison with that in the case of
the signals, generated by the monoenergetic $^7$Be and pep neutrinos,
is not difficult to understand qualitatively. The Super Kamiokande and
SNO experiments will be sensitive only to $^8$B neutrinos having
relatively high energies (E $\ge 5$ MeV and E $\ge 6.44$ MeV) which
exceed at least by a factor of 6 the energy of the dominant (0.862
MeV) component of the $^7$Be neutrino flux. For these energies and the
values of $\Delta m^2$ from the intervals (1a) and (1b) one has:
$2\pi(\epsilon {\rm R}_0/{\rm L}_v) \le 0.14$. As it follows from
eqs. (3) and (4), under this condition the seasonal changes of the
probability ${\rm P}(\nu_e\rightarrow\nu_e; {\rm R(t),E})$ are
proportional to, and do not exceed, the ratio $2\pi(\epsilon {\rm
R}_0/{\rm L}_v)$ and, therefore, cannot be large. The integration over
the neutrino energy reduces further the magnitude of the effect of
interest.

   The solid, dotted, dashed, long--dashed, dash--dotted, and
long--dash--dotted lines in Figs. 3a, 3b and 3c (Figs. 3d, 3e and 3f)
represent results for the values of $\Delta m^2$ and $\sin^2 2\theta$
for which Figs. 1a, 1b,..., 1f (Figs. 2a, 2b,..., 2f) have been
obtained. The normalization of the signals shown graphically in
Figs. 3 is chosen in such a way as to avoid any dependence on the
prediction for the total flux of $^8$B neutrinos, and thus on the
solar models. Namely, for given $\Delta m^2$ and $\sin^2 2\theta$ the
calculated event rate at time t of the year in the case of vacuum
oscillations, $d{\rm N}_{ev}(\Delta m^2,\theta, {\rm E}_{th}; {\rm
t})/dt$, is divided by the quantity $d{\rm N}_{ev}^{0}({\rm t})/dt =
{\rm N}_{ev}(\Delta m^2,\theta,{\rm E}_{th}; {\rm 1 y})~{\rm T}^{-1}
({\rm R}_{0}/{\rm R(t)})^2$, where ${\rm N}_{ev}(\Delta
m^2,\theta,{\rm E}_{th}; {\rm 1 y})$ is the predicted total number of
events per year provided $^8$B neutrinos undergo vacuum oscillations
with the chosen values of the parameters $\Delta m^2$ and $\sin^2
2\theta$, and the ratio $${\rm R}_{var}(\Delta m^2,\theta,{\rm
E}_{th};{\rm t}) = {d{\rm N}_{ev}(\Delta m^2,\theta, {\rm E}_{th};
{\rm t})/dt\over d{\rm N}_{ev}^{0}({\rm t})/dt} = {\rm T}~{{\rm
R^{2}(t)}\over {\rm R}_{0}^{2}}~ {d{\rm N}_{ev}(\Delta m^2,\theta,
{\rm E}_{th}; {\rm t})/dt
\over {\rm N}_{ev}(\Delta m^2,\theta,{\rm E}_{th}; {\rm 1 y})}~, \eqno(5)$$

\noindent is plotted in Figs. 3. For the SNO detector we have:
$${d{\rm N}_{ev}(\Delta m^2,\theta,{\rm E}_{th}; {\rm t})\over dt} =
{1\over {\rm R}^{2}({\rm t})}~\int\limits_{{\rm E}_{th}}^{14.4~MeV}
{\rm F^{SSM}n(E)~P(\nu_e\rightarrow\nu_e;R(t),E)~
\sigma (\nu_{e}d \rightarrow e^{-}pp)}~dE,\eqno(6)$$
$${\rm N}_{ev}(\Delta m^2,\theta,{\rm E}_{th}; 1~y) =
\int\limits_{0}^{T}~ {d{\rm N}_{ev}(\Delta m^2,\theta,{\rm E}_{th};
{\rm t})\over dt}~dt,\eqno(7)$$

\noindent where ${\rm F^{SSM}/R^{2}(t)}$ is the predicted total flux
of $^{8}$B neutrinos at the Earth surface at time t of the year, n(E)
is the normalized to 1 spectrum of $^{8}$B neutrinos,
$\int\limits_{0}^{14.4~{\rm MeV}}{\rm n(E)}dE = 1$, E$_{th} =
6.44~$MeV, and $\sigma (\nu_{e}d \rightarrow e^{-}pp)$ is the
cross--section of the charged current reaction $\nu_{e} + d
\rightarrow e^{-} + p + p$ by which the solar neutrinos will be
detected in the SNO experiment. Obviously, expression (6) corresponds
to ideal detection conditions; for the comparison of the theoretical
predictions with the future SNO data it has to be modified by taking
into account the neutrino energy resolution function, the detection
efficiency, etc. of the SNO detector. The expression for the predicted
event rate in the Super Kamiokande detector can be obtained from eq.
(6) by replacing $\sigma (\nu_{e}d \rightarrow e^{-}pp)$ with the
cross-section $\sigma (\nu_{e}e^{-} \rightarrow \nu_{e}e^{-})$ of the
reaction $\nu_{e} + e^{-} \rightarrow \nu_{e} + e^{-}$, and by using
an appropriate value for E$_{th}$; in the case of $\nu_{e}
\leftrightarrow \nu_{\mu (\tau)}$ oscillations the probability ${\rm
P(\nu_e\rightarrow\nu_e; R(t),E)}$ must be substituted with $${\rm
r}_{\nu} + (1 - {\rm r}_{\nu})~{\rm P(\nu_e\rightarrow\nu_e; R(t),E)},
\eqno(8)$$ where ${\rm r}_{\nu} = \sigma (\nu_{\mu}e^{-} \rightarrow
\nu_{\mu}e^{-})/
\sigma (\nu_{e}e^{-} \rightarrow \nu_{e}e^{-}) \cong {1\over 6}$.

   It is not difficult to convince oneself that $d{\rm
N(t)}_{ev}^{0}/dt$ is the event rate at time t of the year if the
total number of events per year is ${\rm N}_{ev}(\Delta m^2,\theta,
{\rm E}_{th}; 1~y)$ and the relevant signal does not exhibit any
additional time dependence, except for the standard R$^{-2}$(t)
geometrical one. The ratio (5), evidently, is independent of the value
of ${\rm F^{SSM}}$ and thus is solar model independent. The comparison
of the predictions presented graphically in Figs. 3 with the data will
be straightforward: as input one needs only the experimentally
measured mean event rate for a given interval of time (one month,
say), and the total number of events observed per year; the latter
will provide the value of ${\rm N}_{ev}(\Delta m^2,\theta, {\rm
E}_{th}; 1~y)$.  All the other quantities entering into the ratio (5),
T, R$_{0}$ and R(t), are known with a high precision. In the absence
of vacuum oscillations the ratios of signals (5) plotted in Figs. 3
will be equal to 1.

     We shall obtain next an approximate but sufficiently accurate and
rather simple analytic expression for the time variation observable
(5), exhibiting its time dependence explicitly. For $\epsilon = 0.0167
\ll 1$ and $2\pi (\epsilon {\rm R}_{0}/{\rm L}_{v}) \leq 0.14 \ll 1$,
the quantity $2\pi (\epsilon {\rm R}_{0}/{\rm L}_{v})\cos(2\pi {\rm
t/T})$ entering into the formula for the probability ${\rm
P(\nu_e\rightarrow\nu_e; R(t),E)}$, can be used as a small expansion
parameter together with $\epsilon \cos (2\pi {{\rm t\over
T}})$. Expressing the oscillating term in ${\rm
P(\nu_e\rightarrow\nu_e; R(t),E)}$ as a power series in $2\pi
(\epsilon {\rm R}_{0}/{\rm L}_{v})\cos(2\pi {\rm t/T})$, and R(t) as a
power series in $\epsilon \cos (2\pi {{\rm t\over T}})$, it is easy to
show that the leading correction in ${\rm N}_{ev}(\Delta m^2,\theta,
{\rm E}_{th}; 1~y)$ due to the ellipticity $\epsilon$ is proportional
to $\epsilon^2$ and does not exceed $5\times10^{-3}$. Thus, up to
corrections $\sim 5\times10^{-3}$, the quantity ${\rm N}_{ev}(\Delta
m^2,\theta,{\rm E}_{th}; 1~y)$ does not depend on $\epsilon$ and can
be obtained by setting $\epsilon$ to 0 in eqs. (6) and (7).  Using
this fact, and keeping in (6) only the terms up to the second order in
$2\pi (\epsilon {\rm R}_{0}/{\rm L}_{v})\cos(2\pi {\rm t/T})$ in the
expansion of ${\rm P(\nu_e\rightarrow\nu_e; R(t),E)}$, one arrives at
the following result for the observable ${\rm R_{var}}(\Delta
m^2,\theta,{\rm E}_{th}; {\rm t})$ for the SNO detector: $${\rm
R_{var}^{SNO}(\Delta m^2,\theta,E_{th};t}) = 1 + \epsilon \cos
(2\pi{{\rm t\over T}})~\sin^2 2\theta~{\rm K^{SNO}(\Delta m^2,
\theta,{\rm E}_{th})},\eqno(9)$$
where $${\rm K^{SNO}(\Delta m^2,\theta,E_{th})} = {\int\limits_{{\rm
E}_{th}}^{{\rm 14.4~MeV}}dE~{\rm x(\sin 2x - x\cos 2x~\epsilon\cos
2\pi{t\over T})~ n(E)~\sigma (\nu_{e}d \rightarrow e^{-}pp)}
\over \int\limits_{{\rm E}_{th}}^{14.4~MeV}dE~{\rm P(\nu_e\rightarrow\nu_e;
{\rm R_{0},E})~n(E)~\sigma (\nu_{e}d \rightarrow e^{-}pp)}} +
O((2x)^3), \eqno(10)$$ and ${\rm x = \epsilon\pi R_{0}/L_{v}} \le
0.07$.  The corresponding expression for the time variation observable
for the Super Kamiokande detector, ${\rm R_{var}^{SK}(\Delta
m^2,\theta, E_{th};t)}$, can be obtained formally from eqs. (9) and (10)
by replacing in eq. (10) the probability ${\rm
P(\nu_e\rightarrow\nu_e; {\rm R(t),E})}$ by $[{\rm r}_{\nu} + (1 -
{\rm r}_{\nu})~{\rm P(\nu_e\rightarrow\nu_e;R(t),E)]}$, $\sigma
(\nu_{e}d \rightarrow e^{-}pp)$ with $\sigma (\nu_{e}e^{-} \rightarrow
\nu_{e}e^{-})$, by changing the value of ${\rm E_{th}}$, and by
multiplying the numerator in eq. (10) by the factor $(1 - {\rm
r}_{\nu})$. Let us note that the effect of the time dependence of
${\rm K^{SNO(SK)}(\Delta m^2,\theta,E_{th})}$ on ${\rm
R_{var}^{SNO(SK)}(\Delta m^2,\theta,E_{th};t)}$ is beyond the
sensitivity of the next generation of experiments.

   Few comments concerning the results shown in Figs. 3 are in
order. All (correspondingly normalized) signals are equal to 1 at t =
${1\over4}$T and t = ${3\over 4}$T, in accordance with eqs. (9) and
(10).  As Figs. 3 indicate, the predicted amplitude of the time
variations of the signal in the SNO detector is typically (but not
always, e.g., compare the dashed and the long--dashed lines in
Figs. 3a and 3c) larger than that in the Super Kamiokande
detector. The difference in the magnitude of the signal time
variations in the two detectors is a consequence of i) the difference
in the minimal $^{8}$B neutrino energy the two detectors are planned
to be sensitive to (6.44 MeV and 5 MeV), ii) the specific neutrino
energy dependence of ${\rm P(\nu_e\rightarrow\nu_e; R(t),E)}$ in
December and June, iii) the difference in the E--dependence of the
cross--sections $\sigma (\nu_{e}d \rightarrow e^{-}pp)$ and $\sigma
(\nu_{e}e^{-} \rightarrow \nu_{e}e^{-})$, and in the case of $\nu_e
\leftrightarrow \nu_{\mu (\tau)}$ oscillations iv) the probability
${\rm P(\nu_e\rightarrow\nu_e; R(t),E)}$ entering into the expression
for the predicted signals in SNO and Super Kamiokande detectors with
different coefficients (see eqs. (6) and (8)): 1 and approximately
${5\over 6}$, respectively. These differences can lead even to a
strong anticorrelation between the signals in SNO and Super Kamiokande
experiments, as in the case of $\Delta m^2 = 6.9\times 10^{-11}~{\rm
eV}^2$ and $\sin^2 2\theta = 0.9$ (the solid lines in Figs. 3d and
3f).

  If $|{\rm K^{SNO(SK)}(\Delta m^2,\theta,E_{th})}| \ll 1$ for certain
values of $\Delta m^2$ and $\sin^2 2\theta$, one has ${\rm
R_{var}^{SNO(SK)}(\Delta m^2,\theta,E_{th};t)} = 1 + 0(10^{-3})$, and
the time variation effects will not be observable in SNO (Super
Kamiokande) experiment in spite of vacuum oscillations of $^{8}$B
neutrinos. Such is practically the case with the signal in the Super
Kamiokande detector for E$_{th} = 5~$MeV, $\Delta m^2 = 6.3\times
10^{-11}~{\rm eV}^2$ and $\sin^2 2\theta = 0.85$ (see Figs. 1e and 2e
and the dash--dotted lines in Figs. 3a and 3d).  Fortunately, our
results show that the indicated possibility is never realized both for
the signals in the Super Kamiokande and the SNO detectors (compare the
dash--dotted lines in Figs. 3a, 3d and in Figs. 3c and 3e).  Moreover,
it can take place either for the event rate in the Super Kamiokande
detector measured with E$_{th} = 5~$MeV, or for the event rate
obtained with E$_{th} = 7.5~$MeV (the long--dashed line in Fig. 3b),
but not for both event rates (compare the dash--dotted lines in
Figs. 3a, 3d and in Figs. 3b, 3e, and the long--dashed lines in
Figs. 3b and 3a).

    In certain cases the magnitude and the pattern of the time
variation of the signal in the Super Kamiokande detector is very
sensitive to the increase of the threshold neutrino energy from
E$_{th} = 5~$MeV to E$_{th} = (7-8)~$MeV. This is illustrated in
Figs. 3b and 3e, where the results of the calculations of the ratio of
signals (5) for the same values of $\Delta m^2$ and $\sin^2 2\theta$,
for which Figs. 3a and 3d were obtained, but with E$_{th} = 7.5~$MeV
(instead of E$_{th} = 5~$MeV), are presented. We see, in particular,
that for $\Delta m^2 = 6.9\times 10^{-11}~{\rm eV}^2$ and $\sin^2
2\theta = 0.9$ the pattern of the time variations has changed
completely with the change of E$_{th}$: the maximum of the ratio (5)
is now in December rather than in June, and the predicted variations
of the signals in the SNO and Super Kamiokande detectors are
correlated (rather than anticorrelated). Depending on the value of
$\Delta m^2$ (and $\sin^2 2\theta$), the change of E$_{th}$ from 5 MeV
to 7.5 MeV can increase, or diminish the amplitude of the variations
(compare, e.g., the solid, dotted, dash--dotted, and
long--dash--dotted lines in Figs. 3a and 3b, as well as the dashed and
long--dashed lines in the same two figures); for some values of
$\Delta m^2$ and $\sin^2 2\theta$ the increase is quite
substantial. We have not studied the effect of change of E$_{th}$ on
the time variation of the signal in the SNO detector. However, one can
expect on the basis of the above results that for certain values of
$\Delta m^2$ and $\sin^2 2\theta$ it can be dramatic.

    It is interesting to note also [14,27] that for certain values of
the parameters $\Delta m^2$ and $\sin^22\theta$ the seasonal change of
the $^8$B neutrino induced signals, associated with the vacuum
oscillations, can compensate partially or completely the standard 7\%
geometrical one and in the second case the event rate $d{\rm
N_{ev}(\Delta m^2,\theta,E_{th};t)} /dt$ will be constant in time (see
Figs. 1a--1d and 2a and the corresponding solid, dotted, dashed and
long--dashed lines in Fig. 3a, the dotted line in Fig. 3b, the dashed
and long--dashed lines in Fig.  3c, as well as the solid line in
Fig. 3d); it can even lead to an increase of the event rate $d{\rm
N}_{ev}(\Delta m^2,\theta, {\rm E}_{th}; {\rm t})/dt$ from December to
June \footnote{In the case of the monoenergetic $^7$Be and pep
neutrinos even a dramatic increase of the corresponding signals from
December to June due to the vacuum oscillations is possible (as can be
seen in Figs. 1 and 2).}  (see the solid line in Fig. 3b and the solid
and dotted lines in Fig. 3c).  Note that due to the specific
normalization chosen by us a constant event rate will correspond to an
increase of the ratio (5) plotted in Figs. 3 from the value $(1 -
2\epsilon)$ in December to the value $(1 + 2\epsilon)$ in June; an
increase of the event rate from December to June corresponds to an
increase of the ratio (5) from a value smaller than $(1 - 2\epsilon)$
in December to a value greater than $(1 + 2\epsilon)$ in June (see the
solid line in Fig. 3b and the solid and dotted lines in
Figs. 3c). Thus, a non-observation of the 7\% change of the $^{8}$B
neutrino induced event rate (constant rate), or a registration of an
increase of the rate, in the period from December to June in SNO
and/or Super Kamiokande detector would be a strong evidence that solar
neutrinos take part in vacuum oscillations. Note that, as is clear
from Figs. 3, for given $\Delta m^2$ and $\sin^22\theta$ the
compensation (partial or complete) of the standard 7\% seasonal
variation can take place either for SNO or for Super Kamiokande
signals, but not for the signals in both detectors. Futhermore, in the
case of the signal in the Super Kamiokande detector such a
compensation does not hold both for E$_{th} = 5~$ MeV and for E$_{th}
= 7.5~$MeV.
\vskip 0.4cm
\leftline{\bf 2.2 Spectra Deformations}
\vskip 0.2cm
    If solar neutrinos take part in vacuum oscillations, the shapes of
the spectra of the $^{8}$B, pp, and the CNO neutrino fluxes at the
Earth surface will differ from their standard forms. The
corresponding spectra deformations will reflect the specific and
relatively strong dependence of the oscillation probability
${\rm P(\nu_e\rightarrow\nu_e; R(t),E)}$, eq. (3), on the neutrino energy
E. The change of the solar neutrino spectrum will lead also to a
change in the spectrum of the final state electrons in the $\nu
-~$e$^{-}$ elastic scattering reaction induced by the solar neutrinos.

    The deformation of the (average) spectrum of $^{8}$B neutrinos
\footnote{The spectra under discussion will also exhibit relatively small
seasonal variations if solar neutrinos undergo vacuum
oscillations. Here we have in mind the average spectrum which will be
determined experimentally from data collected during a period of k
years, k = 1,2,3,... . The relative magnitude of the correction due to
the seasonal variations in the average spectrum of $^{8}$B neutrinos
is not greater than $\sim 5\times 10^{-3}$, while the relative
difference between the spectra in December and June does not exceed
14\%.} for the same 12 values of the parameters
$\Delta m^2$ and $\sin^22\theta$, for
which Figs.  1a--1f, 3a--3c and Figs. 2a--2f, 3d--3f have been
obtained, are shown respectively in Fig. 4a and Fig. 4b. Each
(average) spectrum, $d{\rm \Phi_{B}(\Delta m^2,\theta,E)}/dE$, to be determined
from data collected by the SNO experiment over a period of k years,
$${d{\rm \Phi_{B}(\Delta m^2,\theta,E)}\over dE} = {1\over {\rm
kT}}~{1\over \sigma (\nu_{e}d \rightarrow e^{-}pp)}~ {d{\rm
N_{ev}(\Delta m^2,\theta,E; k)}\over dE} =$$
\vskip -0.6truecm
\hskip 4.5truecm $$ = {\rm {F^{SSM}\over R_{0}^{2}}~n(E)~
      P(\nu_e\rightarrow\nu_e;R_0,E)}, \eqno(11)$$

\noindent $d{\rm N_{ev}(\Delta m^2,\theta,E; k)}/dE$ being the total
number of events induced by $^{8}$B neutrinos with energy E in
k=1,2,3,...  years, while the last term in eq. (11) represents the
theoretical expression for the spectrum in the case of vacuum
oscillations, is divided by the (average) SSM spectrum, $$ {d{\rm
\Phi_{B}^{SSM}(E)}\over dE} = {\rm {F^{SSM}\over R_{0}^{2}}~n(E)},
\eqno(12)$$ predicted in the absence of oscillations. To avoid the
dependence on the SSM prediction for the total flux of $^{8}$B
neutrinos this ratio of spectra, $${\rm R_{sp}^{SNO}(\Delta
m^2,\theta,E)} = {d{\rm \Phi_{B}(\Delta m^2,\theta,E)}/dE\over d{\rm
\Phi_{B}^{SSM}(E)}/dE}, \eqno(13)$$ is further normalized to the value
of the ratio at E = 10 MeV, and the double ratio $${{\rm
R_{sp}^{SNO}(\Delta m^2,\theta,E)\over R_{sp}^{SNO}(\Delta
m^2,\theta,10~MeV)}} = {{\rm (n(E)~\sigma (\nu_{e}d \rightarrow
e^{-}pp))_{E=10~MeV}\over n(E)~\sigma (\nu_{e}d \rightarrow
e^{-}pp)}}~ {d{\rm N_{ev}(\Delta m^2,\theta,E; k)}/dE\over d{\rm
N_{ev}(\Delta m^2,\theta,10~MeV; k)}/dE} =$$
\vskip -0.6truecm
\hskip 4.0truecm $$ = {{\rm P(\nu_e\rightarrow\nu_e;R_0,E)\over
P(\nu_e\rightarrow
 \nu_e;R_0,E=10~MeV)}},\eqno(14)$$

\noindent is plotted in Figs. 4a and 4b. Thus, in the case of
absence of deformations the ratio of spectra depicted will be constant
(i.e., neutrino energy independent) and equal to 1
\footnote{The absolute deformations of the spectra of the $^{8}$B and pp
neutrinos in the case of $\nu_e\leftrightarrow\nu_{\mu(\tau)}$ (or
$\nu_e\leftrightarrow\nu_s$) oscillations, and for the SSM predictions
of ref. [5] have been shown in ref. [27] for four pairs of values of
the parameters $\Delta m^2$ and $\sin^22\theta$, namely, for ($\Delta
m^2$[eV$^2$]; $\sin^22\theta$) = (1.1$\times10^{-10}$; 1.0),
(7.9$\times10^{-11}$; 0.8), (6.3$\times10^{-11}$; 0.8),
(5.5$\times10^{-11}$; 1.0).}. Note that this
would be valid both for a constant reduction of the spectrum of the
flux (and therefore of the total flux) of $^{8}$B neutrinos by a
certain (energy independent) factor, and if there is no reduction at
all and the flux coincides with the predicted one.

     The changes of the (average) spectrum of the final state e$^{-}$
in the $\nu -$e$^{-}$ elastic scattering reaction induced by the
$^{8}$B neutrinos in the cases of
$\nu_e\leftrightarrow\nu_{\mu(\tau)}$
\footnote {More precisely, induced by the "surviving" $^{8}$B electron
neutrinos and by the $\nu_{\mu (\tau)}$ neutrinos into which the
$^{8}$B neutrinos have oscillated.}, and of
$\nu_e\leftrightarrow\nu_s$ oscillations are shown respectively in
Fig. 5a, Fig. 5b (curves labelled 1--4), and in Fig. 5b (curves
labelled 5 and 6), for the same 12 values of $\Delta m^2$ and
$\sin^22\theta$ for which Figs. 1, 2, 3 and Figs. 4a, 4b have been
obtained.  The e$^{-}$ kinetic energy range chosen (5 MeV$~\leq~$T$_e
\aprle~$14 MeV) coincides with the one to which the Super Kamiokande
detector is planned to be sensitive. The recoil-electron spectra
depicted in Figs. 5a and 5b are normalized in the same way as the
spectra shown in Figs. 4a and 4b
\footnote{In ref.
\cite{KP2} (see Fig. 3a) we have shown just the ratio of the e$^{-}$spectrum
in the case of $\nu_e\leftrightarrow\nu_{\mu(\tau)}$ oscillations, and
of the standard e$^{-}$spectrum, for four pairs of values of $\Delta
m^2$ and $\sin^22\theta$, chosen from different parts of the intervals
(1a) and (1b).}, i.e., the following double ratio is plotted in
Figs. 5a and 5b: $${{\rm R_{sp}^{SK}(\Delta m^2,\theta,T_e)\over
R_{sp}^{SK}(\Delta m^2,\theta,10~MeV)}} = {\rm w(T_e)}~ {d{\rm
N_{ev}(\Delta m^2,\theta,T_e; k)}/dT_e\over d{\rm N_{ev}(\Delta
m^2,\theta,10~MeV; k)}/dT_e} = $$
\vskip -0.4truecm
$$ = {\rm w(T_e)}~{{\rm \int\limits_{T_e(1 + {m_e\over 2T_e})}^{14.4~MeV}
n(E)~(r'_{\nu} + (1 - r'_{\nu})P(\nu_e\rightarrow\nu_e;R_{0},E))}~
(d\sigma (\nu_{e}e^{-}\rightarrow\nu_e e^{-})/dT_e)~dE
\over {\rm \int\limits_{10.25~MeV}^{14.4~MeV}
n(E)~(r'_{\nu} + (1 - r'_{\nu})P(\nu_e\rightarrow\nu_e;R_{0},E))}~
(d\sigma (\nu_{e}e^{-}\rightarrow\nu_e e^{-})/dT_e)~dE}~,\eqno(15)$$

\noindent where $d{\rm N_{ev}(\Delta m^2,\theta,T_e; k)}/dT_e$ is the number
of events (observed in k years) with the recoil e$^{-}$ having an
energy T$_e$, $d\sigma (\nu_{e}e^{-}\rightarrow\nu_e e^{-})/dT_e$ is
the differential cross--section of the process $\nu_{e} + e^{-}
\rightarrow \nu_e + e^{-}$, $r'_{\nu} = (d\sigma (\nu_{\mu}e^{-}
\rightarrow \nu_{\mu}e^{-})/dT_e)/ (d\sigma (\nu_{e}e^{-} \rightarrow
\nu_{e}e^{-})/dT_e) \cong ({1\over 6}- {1\over 7})$, and

$${\rm w(T_e)} = {{\rm \int\limits_{10.25~MeV}^{14.4~MeV}
n(E)}~(d\sigma (\nu_{e}e^{-}\rightarrow\nu_e e^{-})/dT_e)~dE
\over {\rm \int\limits_{T_e(1 + {m_e\over 2T_e})}^{14.4~MeV}
n(E)}~(d\sigma (\nu_{e}e^{-}\rightarrow\nu_e e^{-})/dT_e)~dE}~.\eqno(16)$$

\noindent Thus, in the absence of deformations (no
reduction, or energy independent reduction of the $^{8}$B electron
neutrino flux) the double ratio of e$^{-}-$spectra (15) will represent
a horizontal line crossing the vertical axis at the point 1.

    Let us note that one can choose to normalize the ratios of the
predicted and the standard spectra discussed above by their values not
at 10 MeV, but at some other (in general, different for SNO and Super
Kamiokande detectors) energies.  For a given experiment the energy of
normalization must be chosen on the basis of considerations of
accuracy of the corresponding data, and of maximal enhancement of the
effect of deformation if present in the spectrum.

    One can utilize an alternative spectrum normalization based on the
measurement of the total (average) flux of $^{8}$B neutrinos with
energy E$\geq {\rm E}_{th}$ to form a solar model independent
observable. In the case of the SNO detector this total flux is given
by $${\rm \Phi_{B}(\Delta m^2,\theta,E_{th})} = {\rm
\int\limits_{E_{th}}^{14.4~MeV}}~dE~ {d{\rm \Phi_{B}(\Delta
m^2,\theta,E)}\over dE}~,\eqno(17)$$

\noindent where the integrand is determined by eq. (11). In the absence of
vacuum oscillations (or MSW transitions) the spectrum of $^{8}$B
neutrinos having energies E$\geq {\rm E}_{th}$, whose total flux is
${\rm \Phi_{B}(\Delta m^2,\theta,E_{th})}$, will have the form: $d{\rm
\Phi_{B}^{0}(E)}/dE = {\rm n(E)~\Phi_{B}(\Delta m^2,
\theta,E_{th})}$. The total flux ${\rm \Phi_{B}(\Delta m^2,\theta,E_{th})}$
(or the spectrum $d{\rm \Phi_{B}^{0}(E)}/dE$) can be used to normalize
the measured spectrum (11). Thus, instead of the double ratio (14) one
can consider solar model independent ratio
$${\rm {\bar R}_{sp}^{SNO}(\Delta m^2,\theta,E)} =
{1\over {\rm \Phi_{B}(\Delta
m^2,\theta,E_{th})}}~ {d{\rm \Phi_{B}(\Delta m^2,\theta,E)}\over dE}
=$$
\vskip -0.4truecm
$$ = {({\rm \sigma (\nu_{e}d \rightarrow e^{-}pp)})^{-1}~ d{\rm
N_{ev}(\Delta m^2,\theta,E; k)}/dE\over {\rm
\int\limits_{E_{th}}^{14.4~MeV}}dE~(\sigma (\nu_{e}d \rightarrow
e^{-}pp))^{-1}~d{\rm N_{ev}(\Delta m^2,\theta,E; k)}/dE} = $$
\vskip -0.4truecm
$$ = {1\over{\rm n'(\Delta m^2,\theta,E_{th})}}~{\rm n(E)~
       P(\nu_e\rightarrow\nu_e;R_{0},E)},\eqno(18)$$ where $${\rm
       n'(\Delta m^2,\theta,E_{th})} = {\rm \int
\limits_{E_{th}}^{14.4~MeV}n(E)~P(\nu_e
\rightarrow\nu_e;R_{0},E)}~dE~\eqno(19)$$
is the total $^{8}$B neutrino flux suppression factor in the case of
vacuum oscillations, $0 \leq {\rm n'(\Delta m^2,\theta,E_{th})} \leq
1$.  The analogous ratio for the Super Kamiokande detector can be
easily derived. There are two advantages in utilizing the
normalization described above: i) the corresponding ratios of spectra
will de determined experimentally with a higher precision than the
double ratios (14) and (15), and ii) it allows a straightforward
comparison between the theoretical predictions and the data. For
certain values of $\Delta m^2$ and $\sin^2 2\theta$ the spectra
deformations can be less pronounced in the ratios of the type (18)
than in the double ratios (14) and (15), and vice versa. This is
illustrated in Figs. 5a and 5b, where we show the $^{8}$B neutrino
spectra depicted respectively in Figs. 4a and 4b, but normalized in the
manner described above, eq. (18).
\vskip 0.4cm
\leftline{\bf 3. MSW TRANSITIONS: IMPRINTS ON THE SPECTRA}
\vskip 0.3cm
  In the case of two-neutrino MSW transitions
$\nu_e\rightarrow\nu_{\mu(\tau)}$ or $\nu_e\rightarrow\nu_s$ in the Sun, the
solar $\nu_e$ survival probability, P($\nu_e\rightarrow\nu_e;~$E), can
be calculated with very high accuracy for
$\Delta m^2 \aprge 5\times 10^{-8}~$eV$^2$ and $\sin^22\theta \aprge 10^{-3}$
using the simple analytic expression [32,33]:
$${\rm P}(\nu_e\rightarrow
\nu_e;~{\rm E}) = {1\over 2} + \bigl({1\over2} -
{\rm P}^{'}\bigr)\cos2\theta_m({\rm t}_0) \cos2\theta. \eqno(5)$$
Here
$${\rm P}^{'} = {{\exp\bigl[-\pi{\rm r}_0{\dmf\over{2p}}(1 - \cos2\theta)
\bigr] - \exp\bigl[-2\pi{\rm r}_0{\dmf\over{2p}}\bigr]}\over
{1 - \exp\bigl[-2\pi{\rm r}_0{\dmf\over{2p}}\bigr]}}\eqno(6)$$
\vskip 0.2cm
\noindent is the level crossing probability (i.e., the analog of the
Landau--Zener probability) for the case of density varying
exponentially along the neutrino trajectory in the Sun,
$\theta_m$(t$_0$) is the neutrino mixing angle in matter [13] in the
point of $\nu_e$ production in the Sun, and ${\rm r}_0$ is the
"running" scale height [32,33] (see also [14]), i.e., the scale height
calculated at the resonance point. For $\Delta m^2 \aprge 5\times
10^{-8}~$eV$^2$ and $\sin^22\theta \aprge 10^{-3}$ expression (5)
allows one to derive MSW predictions for any observable quantity
associated with the detection of the solar neutrino flux, or of its
different components, on Earth \footnote {A very precise and simple
analytic description of the two-neutrino MSW transitions of solar
neutrinos for $\sin^22\theta
\aprle 10^{-3}$ was derived in ref. [33]. If $\Delta m^2 \aprle
5\times 10^{-8}~$eV$^2$, for $\sin^22\theta \aprge 0.1$ one must take
into account in the description of the transitions of the
monoenergetic $^{7}$Be and pep neutrinos also the nonadiabatic
oscillating term present in P($\nu_e\rightarrow\nu_e$; E) [34], for
which there exists a relatively simple analytic expression as well
[32].}.

    We have re-examined (exploiting the $\chi^2-$method) the MSW
solution of the solar neutrino problem using the most recent published
data from all four operating solar neutrino detectors (see
Fig. 7). The analysis was based on the SSM predictions of ref.
[5]. It revealed that in the case of $\nu_e\rightarrow\nu_{\mu(\tau)}$
transitions i) the "lower" (in values of $\Delta m^2$) branch of the
large mixing angle (adiabatic) solution [14] (actually, the region
$\Delta m^2 < \times 10^{-6}~$eV$^2$, $\sin^22\theta > 0.1$) is
excluded by the current data at 99.5\% C.L., ii) the "upper" branch
[14] provides a not very good quality of the fit of the data
(min$~\chi^2 = 5.30$ (with the theoretical uncertainties included in
the analysis) for 2 d.f.), being excluded at 90\% C.L., but allowed at
95\% C.L., and iii) the small mixing angle (nonadiabatic) solution
[14] provides the best fit of the data (min$~\chi^2 = 0.48$ (with the
theoretical uncertainties included in the analysis) for 2 d.f.). The
results are quite different if one assumes that
$\nu_e\rightarrow\nu_s$ transitions take place: in this case only a
small mixing angle (nonadiabatic) solution is acceptable (at 90\%
C.L.: min$~\chi^2 = 3.43$ (without the inclusion of the theoretical
uncertainties in the analysis) for 2 d.f.), while a large mixing angle
solution is excluded at 99.7\% C.L. Note, however, that the
nonadiabatic solution in the case of $\nu_e\rightarrow\nu_{\mu(\tau)}$
transitions gives a better quality of the fit of the data than the
nonadiabatic solution associated with the $\nu_e\rightarrow\nu_s$
transitions. Our results are presented graphically in Figs. 7, where
the regions of values of $\Delta m^2$ and $\sin^22\theta$, allowed at
90\% C.L. and 95\% C.L. are depicted: Figs. 7a, 7c and Figs. 7b, 7d
correspond respectively to $\nu_e\rightarrow\nu_{\mu(\tau)}$ and
$\nu_e\rightarrow\nu_s$ conversions.

    The distortions of the spectrum of the $^{8}$B neutrinos (E$\geq
5~$ MeV) predicted in the case of two--neutrino MSW transitions are
shown in Figs. 4c, 4d, and 5c, 5d, while the corresponding distortions
of the spectrum of e$^{-}$ from the reaction $\nu + e^{-} \rightarrow
\nu + e^{-}$ induced by the $^{8}$B neutrinos are depicted in Figs. 6c
and 6d.  The MSW spectra shown in Figs. 4c (5c), 4d (5d) and 6c, 6d
are normalized in the same way as the vacuum oscillation spectra
depicted in Figs. 4a (5a), 4b (5b) and 6a, 6b. Thus, plotted in Figs.
4c (5c), 4d (5d) and 6c, 6d are the corresponding double ratios (14)
and (15) (ratio (18)).  In previous publications we have shown
graphically just the ratio of the predicted MSW and the standard
e$^{-}-$spectra (ref. [14], Fig. 3b), and the absolute deformations of
the $^{8}$B and pp neutrino spectra (ref.  [27], Fig. 5d) for the same
four pairs of values of $\Delta m^2$ and $\sin^22\theta$, for which
Figs. 4c and 6c are obtained. As is evident from Figs. 4--6, the measurements
of the $^{8}$B neutrino and of the recoil--electron spectra in SNO, Super
Kamiokande and ICARUS experiments will allow one, in particular, to
discriminate
between the MSW nonadiabatic and the MSW adiabatic solutions of the solar
neutrino problem.
\vskip 0.4cm
\leftline{\bf 4. DISTINGUISHING BETWEEN THE VACUUM OSCILLATION AND}
{\bf THE MSW SOLUTIONS}
\vskip 0.3cm
    An unambiguous evidence of vacuum oscillations of solar neutrinos
would be the observation of clear deviations from the standard 7\%
seasonal variation of the signals in the future solar neutrino
detectors: no other solution of the solar neutrino problem leads to
such an effect. In the case of vacuum oscillations the predicted
nonstandard seasonal changes of the signals due to the monoenergetic
$^{7}$Be and pep neutrinos are the most dramatic (see Figs. 1 and
2). Although much smaller, the seasonal variation effects in the
signals generated by the $^{8}$B neutrinos are, for most of the values
of $\Delta m^2$ and $\sin^22\theta$ from the intervals (1a) and (1b),
sufficiently large to be detected by the SNO, Super Kamiokande and
ICARUS experiments, provided the detectors will operate with their
envisaged detection capabilities and expected background levels. As we
have demonstrated, the effects can be enhanced by choosing appropriate
values of the relevant threshold detection energies. The data on the
seasonal time variations of the event rates in SNO, Super Kamiokande
and ICARUS experiments can be crucial for discriminating between the
vacuum oscillation and the other possible solutions of the solar
neutrino problem.

      The predicted distortions of the $^{8}$B neutrino and the
recoil--electron spectra due to two--neutrino vacuum oscillations or
MSW transitions of the $^{8}$B neutrinos (Figs. 4, 5 and 6) provide us
with an indispensible possibility to test these solutions in a solar
model independent way in SNO, Super Kamiokande and ICARUS
experiments. As is evident from the comparison of Figs. 4a,...,4d
(5a,...,5d) and Figs. 6a,...,6d, respectively, both the vacuum
oscillations and the MSW transitions lead to somewhat stronger shape
deformations of the $^{8}$B neutrino spectrum than of the
recoil--electron spectrum: some of the features of the distorted
$^{8}$B neutrino spectrum are less pronounced, or are not present, in
the e$^{-}$ spectrum as a result of the integration over the neutrino
energy necessary to perform to obtain the latter. The only exception
are the spectra corresponding to large mixing angle MSW transitions
(see the curves labelled 2 and 3 (4 and 5) in Figs. 4c, 5c and 6c
(Figs. 4d, 5d and 6d)). The predicted spectra deformations in this
case are rather small and, most probably, will be difficult to detect
in SNO, Super Kamiokande and ICARUS experiments.  Let us add that the
distortions of the spectra can be enhanced by an appropriate
choice of the specific normalization of the spectra, with the help of
which one forms solar model independent spectrum observables (as a
comparison of Figs. 4 and 5 indicates). The results depicted in
Figs. 4 and 5 show that the vacuum oscillations, the MSW adiabatic,
and the MSW nonadiabatic transitions of solar neutrinos lead to
distinctly different deformations of the spectrum of the $^{8}$B
neutrinos, to be measured in the SNO (ICARUS) experiment. It seems
very likely that the data from the SNO (ICARUS) detector on the
$^{8}$B neutrino spectrum will allow one to test and to discriminate
between these three possibilities. Adding the information about the
seasonal variations of the signal will, most probably, permit to
unambiguously distinguish between the vacuum oscillation and the MSW
solutions.

    The distortions of the recoil--electron spectrum shown in Figs. 6
suggest that on the basis of the Super Kamiokande (ICARUS) data on the
e$^{-}$ spectrum alone it may be difficult to discriminate between
vacuum oscillations with $4.5\times 10^{-11} {\rm eV}^2 \aprle \Delta
m^2 \aprle 6.3\times10^{-11} {\rm eV}^2$ and MSW nonadiabatic
transitions of solar neutrinos (compare curves 5 and 6 in Fig. 6a, and
2--5 in Fig. 6b with curves 4 in Fig. 6c and 1--3 in Fig.  6d). In the
case of vacuum oscillations with $\Delta m^2$ from the above interval
there will be seasonal variations of the signals in the Super
Kamiokande, SNO and ICARUS detectors (see the dotted, dashed and
long--dashed curves in Figs. 3d, 3e and 3f), which can be used to
eliminate one of these two possibilities.

    It is also clear from Figs. 4, 5 and 6 that the information about
the shapes of the $^{8}$B neutrino and the e$^{-}$ spectra to be
obtained in the SNO, Super Kamiokande and ICARUS experiments, most
probably, will not be sufficient to discriminate between an
astrophysical and the large mixing angle (adiabatic) MSW solutions of
the solar neutrino problem. However, the measurement of the ratio of
the number of events due to the solar neutrino induced charged current
(CC) and neutral current (NC) reactions on deuterium, ${\rm
R}^{CC/NC}$, to be performed with a relatively high precision in SNO
experiment, will provide a crucial test of the large mixing angle MSW
solution: for this solution one has ${\rm R}^{CC/NC}_{AS} \cong (0.3 -
0.4)~{\rm R}^{CC/NC}$, where ${\rm R}^{CC/NC}$ is the value of the
ratio predicted in the absence of oscillations and/or of MSW
transitions. Note that the quantity ${\rm R}^{CC/NC}$ does not depend
on the total flux of $^{8}$B neutrinos, and therefore is solar model
independent; the value of ${\rm R}^{CC/NC}$ can be calculated with a
high precision.

    To conclude, the envisaged capabilities of the next generation of
solar neutrino experiments will allow one to perform crucial solar
model independent tests of, and to discriminate between, the vacuum
oscillation and the MSW solutions of the solar neutrino problem. It is
very likely that the "solar neutrino puzzle" will be resolved by these
experiments.
\vskip 0.4cm
\leftline{\bf Acknowledgements.} The kind hospitality and partial support of
the Institute for Nuclear Theory at the University of Washington in
Seattle, where part of the present work was done, is acknowledged with
gratefulness.  The work of P.I.K. has been supported by grant
No. DE--FG05--85ER--40219 of the U.S. Department of Energy and by a
grant from the North Carolina Supercomputing Program.

\newpage
\centerline{\bf Figure Captions}
\medskip
\noindent
{\bf Fig. 1} The ratio of the vacuum oscillation
($\nu_e\leftrightarrow\nu_{\mu(\tau)}$) and of the SSM predicted
signals (event rates) due to the $^{8}$B, pep, $^{7}$Be, and pp
neutrinos as a function of the time of the year (in units of 365
days). The results shown are for experiments detecting the solar
neutrinos via the $\nu$ -- e$^{-}$ elastic scattering reaction (Super
Kamiokande, BOREXINO, HELLAZ, etc.). The SSM predicted signals used
represent the time independent one year average values of the event
rates calculated within the model [5].

\medskip

\noindent
{\bf Fig. 2} The same as in Fig. 1 for different sets of values of
$\Delta m^2$ and $\sin^2 2\theta$. Figures e and f correspond to
$\nu_e\leftrightarrow\nu_{s}$ oscillations.

\medskip

\noindent
{\bf Fig. 3} Time variations of the signals in Super Kamiokande
(Figs. 3a, 3b, 3d and 3e) and SNO (Figs. 3d and 3f) detectors in the
case of vacuum oscillations of $^{8}$B neutrinos, for the same values
of the parameters $\Delta m^2$ and $\sin^2 2\theta$ for which
Figs. 1a, 1b,..., 1f and Figs.  2a, 2b,..., 2f have been obtained
(solid, dotted, dashed, long--dashed, dash--dotted and
long--dash--dotted lines in Figs. 3a--3c and 3d--3f,
respectively). The signals are normalized in such a way that in the
absence of deviations from the standard 7\% seasonal variation they
will be constant in time and equal to 1 (i.e., horizontal lines
crossing the vertical axis at the point 1); the normalization used
renders solar model independent the ratio of signals plotted. The
results presented in Fig. 3a (3d) and Fig. 3b (3e) have been obtained
with different values of the neutrino threshold energy: E$_{th} =
5~$MeV and E$_{th} = 7.5~$MeV, respectively.

\medskip

\noindent
{\bf Fig. 4} Deformations of the $^{8}$B neutrino spectrum in the
cases of $\nu_e\leftrightarrow\nu_{\mu(\tau)}$ (or
$\nu_e\leftrightarrow\nu_{s}$) oscillations (a and b), and of $\nu_e
\rightarrow \nu_{\mu(\tau)}$ (or $\nu_e\rightarrow\nu_{s}$) MSW
transitions (c and d). The vacuum oscillation and the MSW spectra are
divided by the SSM predicted spectrum [5], and each ratio of spectra
is further normalized to the value this ratio has at E = 10
MeV. The double ratios plotted are solar model independent quantities.

\medskip

\noindent
{\bf Fig. 5} The same as in Fig. 4, but with different normalization
of the spectra (see eq. (18)).

\medskip

\noindent
{\bf Fig. 6} Deformations of the spectrum of e$^{-}$ from the reaction
$\nu + e^{-} \rightarrow \nu + e^{-}$ caused by $^{8}$B neutrinos, in
the cases of oscillations in vacuum
$\nu_e\leftrightarrow\nu_{\mu(\tau)}$ (a, and b (curves 1--4)),
$\nu_e\leftrightarrow\nu_{s}$ (b (curves 5 and 6)), and of MSW
transitions $\nu_e\rightarrow\nu_{\mu(\tau)}$ (c and d). Each of the
predicted recoil--electron spectrum is divided by the standard
one and the ratio so obtained is normalized to the value it has at
T$_{e}=$ 10 MeV.

\medskip
\noindent
{\bf Fig. 7} Regions of values of the parameters $\Delta m^2$ and
$\sin^2 2\theta$ allowed at 90\% C.L. (dashed lines) and at 95\% C.L.
(solid lines) by the current solar neutrino data in the case of MSW
$\nu_e\rightarrow\nu_{\mu(\tau)}$ (a and c) and
$\nu_e\rightarrow\nu_{s}$ (b and d) transitions of solar
neutrinos. Figures a and b (c and d) have been obtained by including
(without including) the uncertainties in the theoretical predictions
[5] in the relevant $\chi^2-$ analysis.

\end{document}